\documentclass{amsart}

\usepackage{epsfig}
\usepackage{graphicx}
\usepackage{amsmath,amssymb,amsfonts,latexsym}
\usepackage{graphicx}

\usepackage{amssymb}
\usepackage{enumerate, xspace}
\usepackage{dsfont}
\usepackage{afterpage}
\usepackage{changebar}
\usepackage{amsthm}
\usepackage{rotating}

\numberwithin{equation}{section}

\begin{document}

\title[Water molecule dissociation]
  {Polarization induced water molecule dissociation below the first-order electronic-phase transition temperature}
\author[Arulsamy et al.]{Andrew Das Arulsamy$^1$} \email{sadwerdna@gmail.com}

\address{$^1$Condensed Matter Group, Division of Interdisciplinary Science, D403 Puteri Court, No. 1, Jalan 28, Taman Putra, 68000 Ampang, Selangor DE, Malaysia}

\author[]{Zlatko Kregar$^2$}

\address{$^2$Institute of Physics, Bijenicka 46, HR-10000 Zagreb, Croatia}

\author[]{Kristina Eler$\check{\rm s}$i$\check{\rm c}$$^3$}

\address{$^3$Jo$\check{\rm z}$ef Stefan Institute, Jamova cesta 39, SI-1000 Ljubljana, Slovenia}

\author[]{Martina Modic$^3$}
\author[]{Uma Shankar Subramani$^4$}
\address{$^4$MS Software Engineering, Malardalen University, Vasagatan 44, 72123 Vasteras, Sweden}


\date{\today}

\begin{abstract}
Hydrogen produced from the photocatalytic splitting of water is one of the reliable alternatives to replace the polluting fossil and the radioactive nuclear fuels. Here, we provide unequivocal evidence for the existence of blue- and red-shifting O$-$H covalent bonds within a single water molecule adsorbed on MgO surface as a result of asymmetric displacement polarizabilities. The adsorbed H$-$O$-$H on MgO gives rise to one weaker H$-$O bond, while the other O$-$H covalent bond from the same adsorbed water molecule compensates this effect with a stronger bond. The weaker bond (nearest to the surface), the interlayer tunneling electrons and the silver substrate are shown to be the causes for the smallest dissociative activation energy on MgO monolayer. The origin that is responsible to initiate the splitting mechanism is proven to be due to the changes in the polarizability of an adsorbed water molecule, which are further supported by the temperature-dependent static dielectric constant measurements for water below the first-order electronic-phase transition temperature. 
\end{abstract}

\maketitle


Water molecules do not split spontaneously to give hydrogen for clean energy applications, but this is not bad at all because stable water molecules are necessary to sustain life on earth. To split them however, we must apply external disturbances so as to destabilize the water molecules. One elementary way of doing so is via the electrolysis process, in which one needs a relatively high voltage between two electrodes to split the water molecules. Understanding how the splitting process is initiated is extremely important not only to pave the path to obtaining clean energy~\cite{kroto,tour,nam,hema,yer,robin}, but also scientifically because this knowledge can be transferred to our understanding of the origin of chemical reactions~\cite{scr}, beyond the standard quantum mechanical calculations of activation energies~\cite{fang,marco,ohno}. For example, we need to understand the changes to the molecule-oxide interaction strength (that gives rise to efficient water-molecule dissociations) under different external and/or internal disturbances. Varying interaction strengths (due to external or internal disturbances) can be efficiently tracked and evaluated with the ionization energy theory (IET)~\cite{pra} that relies on the energy-level spacing renormalization technique~\cite{ann}.      

Acquiring such knowledge (how water molecules split in the presence of disturbances) require us to go beyond the standard procedure of describing the existence of dissociation energy profile, or the profile of energy levels before and after a particular dissociative reaction path. Such path defines the existence of an activation energy for a given reaction and its thermodynamic process (endothermic or exothermic), but never on why and how this activation energy changes for different interaction strengths. Here, we give unambiguous explanations (i) how to track the changes to the interaction strengths between a water molecule and a given oxide surface, and (ii) why the above interaction strengths vary when exposed to a polarizable ion or surface. The theoretical discussions given here are general and therefore can be applied to other chemical reactions with modifications, depending upon the complexity of the molecules. In a previous report by Shin et al.~\cite{shin}, two types of water molecule dissociation pathways have been selectively achieved by inducing interlayer tunneling electrons through a single water molecule. These dissociations, as well as the water molecule diffusion (due to desorption and hopping) have been observed with the scanning tunneling microscopy (STM). The first dissociation path requires higher energy because it is due to electronic excitations. Whereas, vibrational excitations (for high tunneling current) were found to be responsible for the initiation of low-energy dissociation due to the asymmetric stretching mode, $\nu_{\rm OH}$ ($\approx$ 448 meV)~\cite{shin}. However, at low tunneling current, both the scissoring mode, $\delta_{\rm HOH}$ ($\approx$ 217 meV) and $\nu_{\rm OH}$ contribute to the water-molecule desorption and hopping. 

\begin{figure}
\begin{center}
\scalebox{0.4}{\includegraphics{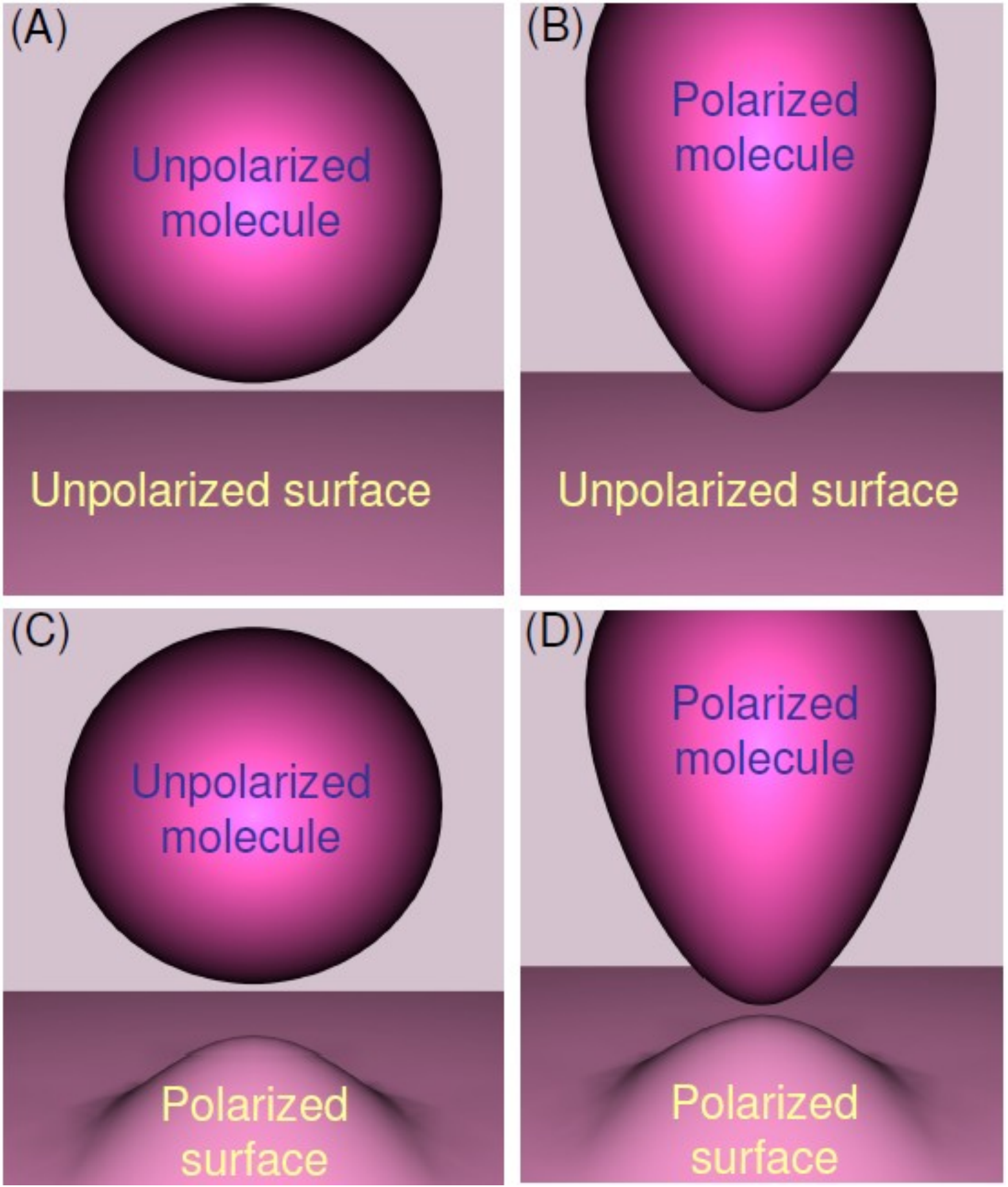}}
\caption{(A) Both the molecules and the oxide surface are unpolarized, and therefore, they do not contribute to physisorption, and with negligible quantum mechanical interaction between them. (B) Shows a polarized molecule on an unpolarized oxide surface. This gives rise to an effective attractive interaction (physisorption) between them. (C) Shows the reverse polarizability compared to (B) that will also contribute to an effective physisorption. (D) Here, both entities are polarized with strong electron-electron ($e$-$e$) repulsion between them (molecule and oxide surface), thus giving rise to an effective desorption and hopping processes. The magnitude of the $e$-$e$ repulsion, size of the molecules and the electron polarization are not to scale. See text for details.}
\label{fig:1}
\end{center}
\end{figure}

There are four types of molecule-oxide interactions (see Fig.~\ref{fig:1}) that can be used to understand the chain of events required to initiate the water molecule dissociation on oxides as stated above. In doing so, we will be able to achieve the motivations listed in (i) and (ii). In addition, we exploit these information to reinforce the accuracy of the Hermansson blue-shifting hydrogen-bond theory~\cite{her1,her2,her3}, and invoke the temperature ($T$)-dependent static dielectric constant measurements~\cite{lide} of water for 0.1, 1, 10 and 100 MPa within IET to prove that our water splitting mechanism is indeed unambigous below the first-order electronic-phase transition temperature. The analyses are entirely analytic and based on the mathematically well-defined ionization energy approximation, which is self-consistent both qualitatively and quantitatively~\cite{pra,ann}. 

First, we note that the electrons from these cations, 2H$^{+}$ and Mg$^{2+}$ determine the polarizability of their respective independent system, H$_2$O molecule and MgO crystal. Their polarizability can be evaluated from their averaged ionization energies, which are given by 1312 and 1094 kJmol$^{-1}$ for 2H$^{+}$ and Mg$^{2+}$, respectively. The averaged atomic ionization energy values for all the elements considered in this work are given in Table~\ref{Table:I}, in which, the averaging follows Ref.~\cite{pla1}. Prior to averaging, all the atomic ionization energies were taken from Ref.~\cite{web}. The polarizability magnitudes can be calculated from~\cite{pla3} 

\begin {eqnarray}
\alpha_{d} = \frac{e^2}{M}\bigg[\frac{\exp[\lambda(E_{\rm F}^0 - \xi)]}{(\omega_{\rm{ph}}^2 - \omega^2)}\bigg]. \label{eq:1}
\end {eqnarray}  

Here, $\omega_{\rm{ph}}$ is the phonon frequency of undeformable ions, $\xi$ denotes the averaged ionization energy, $E^0_{\rm F}$ is the Fermi level when $T$ equals zero (and independent of any disturbance), $1/M = 1/M^+ + 1/M^-$, $M^+$ and $M^-$ are the respective positively and negatively charged ions mass due to their different individual polarizabilities, $\lambda = (12\pi\epsilon_0/e^2)a_B$, in which, $a_B$ is the Bohr radius of atomic hydrogen, $e$ and $\epsilon_0$ are the electron charge and the permittivity of free space, respectively. The above definition for $E^0_{\rm F}$ implies that $E^0_{\rm F}$ can also be associated to the Sachdev critical tuning point, which defines the quantum phase transitions~\cite{sachdev5,sachdev}. For example, one can obtain free-electron metals for electrons with energies $E > E^0_{\rm F}$, while non-free-electron property prevails for $E < E^0_{\rm F}$~\cite{ann}. Equation~(\ref{eq:1}) gives us the direct relationship between the constituent atomic ionization energies (types of ions) and the ionic polarizabilities (with renormalized electronic contribution). 

However, Eq.~(\ref{eq:1}) does not apply for the diffusion of nanoparticles (classical) in the colloidal system~\cite{raj}, where the non-existence of the energy-level spacing and polarizability effects for these nanoparticles have been correctly justified. In the case of IET, the polarizability of one ion, affects its nearest neighbor, and then its next nearest neighbor, and so on depending on the strength of the polarization, which in turn gives rise to the long-range Coulomb interaction. On the other hand, undeformable (charged or uncharged) nanoparticles always satisfy the first-order-short-range interaction that can be treated with classical formalism~\cite{raj,rog}, which is not suitable if the long-range Coulomb interaction dominates~\cite{veh}.

\begin{table}[ht]
\caption{Averaged atomic ionization energies ($\xi$) for individual ions and their respective valence states ordered with increasing atomic number $Z$. All the experimental ionization energy values were obtained from Ref.~\cite{web}.} 
\begin{tabular}{c c c c} 
\hline\hline 
\multicolumn{1}{l}{Elements}           &    Atomic number  &  Valence    & $\xi$   \\  
\multicolumn{1}{l}{}                &   $Z$             &  state      & (kJmol$^{-1}$)\\  
\hline 

H                                   &  1   					    &  1+      & 1312 \\ 
O                                   &  8	   	  			  &  1+      & 1314 \\ 
O                                   &  8	   	  			  &  2+      & 2351 \\ 
Mg                                  &  12	   	  			  &  2+      & 1094 \\ 
 
\hline  
\end{tabular}
\label{Table:I} 
\end{table}

From Eq.~(\ref{eq:1}) and Table~\ref{Table:I}, one finds $\alpha_d^{\rm 2H^+} > \alpha_d^{\rm O^{2+}}$ and $\alpha_d^{\rm Mg^{2+}} > \alpha_d^{\rm O^{2+}}$. Moreover, one can also use Eq.~(\ref{eq:1}) to obtain Fig.~\ref{fig:1} that depicts the importance of water-solid interactions~\cite{thiel}. But let us first derive the above inequalities from the following analysis, which is somewhat subtle, but it will be clear once the electron-transfer between ions are understood. We first need to determine the reason why H acts as a cation, and O as an anion in a H$_2$O molecule within the ionization energy theory. In order to do so, we need to consider the ionization-energy inequality between 2H$^{+}$ and O$^{2+}$, which is given by $\xi_{\rm 2H^{+}} < \xi_{\rm O^{2+}}$ (see Table~\ref{Table:I}). This means that the effective electron-transfer is from H to O atom due to $\alpha_d^{\rm 2H^{+}} > \alpha_d^{\rm O^{2+}}$. Subsequently, we can now extend this argument to claim that the electron-transfer in MgO system is from Mg to O atom due to $\xi_{\rm Mg^{2+}} < \xi_{\rm O^{2+}}$, which implies $\alpha_d^{\rm Mg^{2+}} > \alpha_d^{\rm O^{2+}}$ from Eq.~(\ref{eq:1}) and Table~\ref{Table:I}. As a consequence, the electrons from 2H$^+$ and Mg$^{2+}$ are the ones that will contribute to the polarizabilities of H$_2$O and MgO surface, respectively. 

Here, we generalize the water-MgO surface interaction as follows$-$ if we have a reactive surface with dangling bonds and defects such that the surface electrons are more strongly polarizable (reactive) than the gas molecules, then the stronger interaction between the surface and a molecule leads to an effective physisorption (see Fig.~\ref{fig:1}(B)). However, these polarizability-dependent physisorption mechanism can only be true for a certain narrow temperature-window ($T_1$ to $T_2$, $T_2 > T_1$) or at low temperatures. For $T < T_1$, this attractive interaction remains attractive due to insufficient thermal energy to overcome the strength of physisorption. In contrast, higher temperatures ($T > T_2$) could lead to chemisorption due to chemical reaction (not shown) or desorption (hopping) as depicted in Fig.~\ref{fig:1}(D). Figure~\ref{fig:1}(A) defines the short-range Coulomb interaction as predicted for classical nanoparticles~\cite{raj}. 

Apparently, this narrow window depends on the types of solid surfaces and molecules where (a) larger polarizability of a given molecule will have a stronger physisorption onto the smaller polarizable solid surface, [see Fig.~\ref{fig:1}(B)]. Similarly, (b) larger polarizable solid surfaces (compared to molecules) are also suitable to obtain an effective physisorption [see Fig.~\ref{fig:1}(C)]. On the contrary, (c) unpolarized systems do not interact because the electrons from these two entities have large ionization energies (thus, weaker polarization), for example, atomic inert noble gases do not easily interact at quantum level [see Fig.~\ref{fig:1}(A)]. At the other extreme, (d) two highly polarizable components will give rise to an enhanced desorption, which is undesirable because they can be easily evaporated due to polarized $e$-$e$ repulsion~\cite{cphc} [see Fig.~\ref{fig:1}(D)]. Such a repulsive interaction has been predicted between a single Si ion on the SiC surface in the presence of electric fields. Here, the Si ions (with high polarizability) tend to diffuse and evaporate easily from the SiC substrate compared to C ions (that have small polarizability)~\cite{pla3}. Of course, these four mechanisms are strictly valid only within that narrow temperature range or at low temperatures as pointed out earlier. 

Our focus here however is to exploit the information obtained from Fig.~\ref{fig:1} to determine the physico-chemical properties of a given catalyst needed to split the water molecules efficiently. Apparently, this catalyst is one of the most sought after component for photocatalytic water splitting~\cite{nam,hema,robin,kohl,hei}. Parallel to this, MgO surface on Ag(100) substrate has been investigated for its suitability as a catalyst to split a single water molecule by activating its vibrational and electronic excitations selectively~\cite{shin}. In order to make this vibration-induced dissociation mechanism explicit, we will need to first explain the effect of MgO surface polarizability as a result of Ag substrate underneath. One layer of MgO grown on Ag(100) will have a larger polarizability (on the MgO surface) compared to a two or more MgO layers (we assume defect-free MgO and Ag(100) structures). This larger surface polarizability originates from the electrons contributed by the Ag layer (a free electron metal), in which, MgO is an insulator and thus its (MgO) surface polarizability is much smaller based on IET. For a two-layer MgO on Ag(100), the effect of free electrons from Ag is reduced, giving rise to a lower polarizability of MgO surface than on a single-layer MgO surface~\cite{referee1}. Therefore, one can expect a systematic lowering of the MgO surface polarizability with the increasing number of MgO layers before achieving the bulk MgO polarizability value. Note here that MgO is the least polarizable material compared to Ag.  

\begin{figure}
\begin{center}
\scalebox{0.4}{\includegraphics{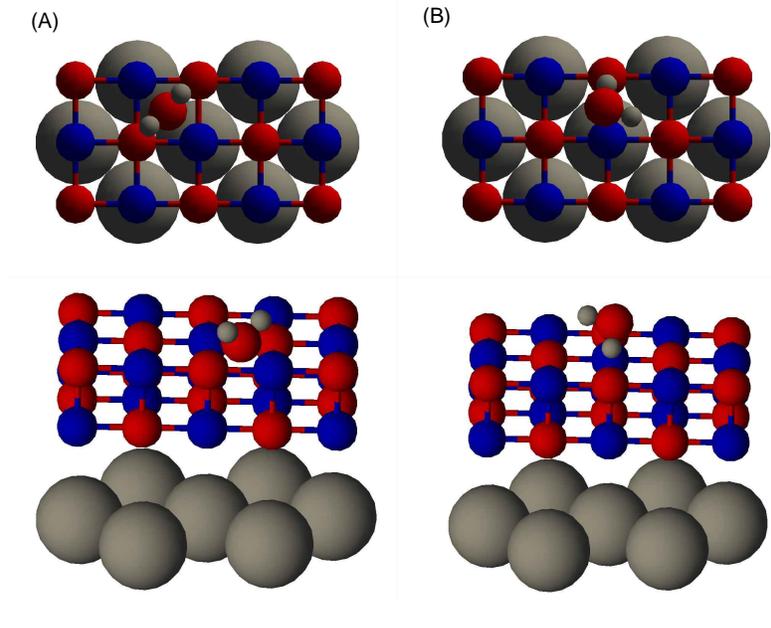}}
\caption{Top view: a single layer of MgO is schematically drawn on top of Ag substrate. Two possible configurations are shown schematically for a single water molecule adsorbed on MgO. (A) The oxygen ion from a water molecule (O$_{\rm WAT}$) is adsorbed on the bridge site, surrounded by two Mg$^{2+}$ and two O$^{\delta -}_{\rm SUR}$ ions (nearest neighbors on the surface). Whereas, the hydrogen ions (H$^{\delta +}_{\rm I}$ and H$^{\delta +}_{\rm II}$) are both dangling upward, and their nearest surface neighbors are O$^{\delta -}_{\rm SUR}$ ions at the bottom. (B) This configuration has been predicted from the DFT calculations~\cite{shin}, in which, H$^{\delta +}_{\rm I}$ is adsorbed onto a O$^{\delta -}_{\rm SUR}$ ion, while H$^{\delta +}_{\rm II}$ ion dangles (not adsorbed) pointing toward another O$^{\delta -}_{\rm SUR}$ ion. In addition, O$^{\delta -}_{\rm WAT}$ ion is also not adsorbed, but its nearest neighbor is the Mg$^{2+}$ ion at the bottom. The ionic sizes are not to scale.}
\label{fig:2}
\end{center}
\end{figure}

Having found the reason for MgO surface polarizability as due to the Ag(100) substrate (free electron metal), we can now go on and develop the mechanism that gives rise to a single water molecule dissociation on one or more MgO layers. There are two possibilities for a H$^{+}_2$O$^{2-}$ molecule to be adsorbed on a Mg$^{2+}$O$^{2-}$ monolayer. One of them is shown in Fig.~\ref{fig:2}A, in which, the oxygen anion (from water) is adsorbed on the bridge site surrounded by two Mg$^{2+}$ and O$_{\rm SUR}$ surface ions such that the two unadsorbed hydrogen ions (from the water molecule) have the same probability to be dissociated. Here, H$_{\rm I}$ and H$_{\rm II}$ ions are dangling upward (not adsorbed) and their nearest surface neighbors are O$_{\rm SUR}$ ions. This configuration is energetically not favorable compared to the one shown in Fig.~\ref{fig:2}B, however, the former configuration is possible in a nonequilibrium condition (with fluctuating energies at nanoscale). One reason to focus on this particular physisorption (Fig.~\ref{fig:2}A) is that it is the only alternative configuration that will give rise to the dissociated product, OH$^{-}$ on the bridge site directly. On the other hand, the physisorption depicted in Fig.~\ref{fig:2}B was predicted to be stable from the density functional theory (DFT) calculations~\cite{shin}, and this configuration requires re-adsorption of OH$^{-}$ on the bridge site, after dissociation. Therefore, it is important to exclude the physisorption shown in Fig.~\ref{fig:2}A as a possibility, which can be done using IET. 

In Fig.~\ref{fig:2}B, the electronic polarization on the surface of MgO monolayer [grown on Ag(100)] is the largest and is expected to gradually decrease with the number of MgO layers (for example, from one to three monolayers) due to the insulating character of MgO. Therefore, one can evaluate Eq.~(\ref{eq:1}) to obtain the largest attractive potential between a positively charged hydrogen ion (H$_{\rm I}^{\delta +}$) and the negatively charged surface oxygen ion (O$_{\rm SUR}^{\rm \delta -}$). Such stronger attractive interaction will cause the covalent bond between H$_{\rm I}^{\delta +}$ and O$_{\rm WAT}^{\rm \delta -}$ (oxygen ion from the adsorbed water molecule) to be weaker (smaller $\nu_{\rm OH}$)~\cite{referee2}. Here, we have used $\rm \delta +$ and $\rm \delta -$ instead of 1+ and 2$-$ for hydrogen and oxygen ions, respectively, so that we can accommodate the changes of these charges as a result of their changing polarizabilities (due to temperature, increasing MgO layers, chemical compositions and defects). The surface-electron polarization induced attractive potential between H$_{\rm I}^{\delta +}$ and O$_{\rm SUR}^{\rm \delta -}$ has a similar effect to the hydrogen bonds between water molecules in liquid or solid state (ice). In particular, this attractive potential should reduce both the symmetric and asymmetric ($\nu_{\rm OH}$) stretching vibrational frequencies for the adsorbed water molecule such that $\nu_{\rm OH}$ value should be lower than the water vapour (isolated water molecules). 

As a consequence, we anticipate $\nu_{\rm OH}$ to range between 454 meV (for water vapour) and 441 meV (for liquid phase water at 313 K) or lower. The above stated values for $\nu_{\rm OH}$ were measured and reported in Ref.~\cite{sow}. Such an inverse proportionality between $\nu_{\rm OH}$ and the polarization induced attractive potential (or the hydrogen bonds between water molecules) arises from the fact that the strength of the charge $\delta +$ from H$_{\rm I}^{\delta +}$ is reduced by the strongly polarized O$_{\rm SUR}^{\rm \delta -}$ as seen by the O$_{\rm WAT}^{\rm \delta -}$. In other words, the stronger attractive Coulomb potential ($V_{\rm Coulomb}^{\rm attract}$) between H$_{\rm I}^{\delta +}$ and O$_{\rm SUR}^{\rm \delta -}$ will reduce the ability of H$_{\rm I}^{\delta +}$ to attract its own (shared) electron from the electron-rich O$_{\rm WAT}^{\rm \delta -}$. Thus, one gets a stronger electric dipole moment ($\textbf{p}$) between H$_{\rm I}^{\delta +}$ and O$_{\rm WAT}^{\rm \delta -}$. The reason for this stronger $\textbf{p}$ can be understood by first noting that the above $V_{\rm Coulomb}^{\rm attract}$ gives rise to smaller $e$-$e$ interaction, for example, between the electrons from O$_{\rm WAT}$ and O$_{\rm SUR}$ through H$_{\rm I}$. This means that H$_{\rm I}$ effectively screens and reduces the $e$-$e$ interaction between those two oxygen ions. This in turn implies a smaller Shankar-type screened Coulomb potential~\cite{shankar}, which can be evaluated from~\cite{pra}

\begin {eqnarray}
\frac{1}{V_{\rm Coulomb}^{\rm attract}} \propto V^{\rm O_{WAT} - O_{SUR}}_{\rm{sc}} = \frac{e}{4\pi\epsilon_0r}\exp{\big[-\mu re^{-\frac{1}{2}\lambda\xi^{\rm O_{WAT}}_{\rm H_{I}}}\big]}, \label{eq:2}
\end {eqnarray}

where $\mu$ is the screening constant of proportionality. Equation~(\ref{eq:2}) indicates that after the physisorption, one obtains a smaller $\xi^{\rm O_{WAT}}_{\rm H_{I}}$ as a result of smaller $V^{\rm O_{WAT} - O_{SUR}}_{\rm{sc}}$. In other words, $\xi^{\rm physisorped}_{\rm O_{WAT}-H_{I}} < \xi^{\rm isolated}_{\rm O_{WAT}-H_{I}}$. Having found the reason why physisorption reduces the ionization energy, $\xi^{\rm O_{WAT}}_{\rm H_{I}}$ for the electron shared between O$_{\rm WAT}$ and H$_{\rm I}$, we can now move on and define        

\begin {eqnarray}
\textbf{p} = e\langle\Psi^{\rm upper}_{\rm state}|\textbf{r}|\Psi^{\rm lower}_{\rm state}\rangle  = \alpha_{d}\textbf{E}^{\rm O_{WAT}}_{\rm H_{I}}. \label{eq:3}
\end {eqnarray}  

Here, $\textbf{r}$ is the electron coordinate, $\Psi^{\rm upper}_{\rm state}$ and $\Psi^{\rm lower}_{\rm state}$ represent the many-body (a water molecule adsorbed on an oxide surface) wave functions for the upper and lower states, respectively, and $\textbf{E}^{\rm O_{WAT}}_{\rm H_{I}}$ denotes the microscopic electric field between O$_{\rm WAT}$ and H$_{\rm I}$. Fortunately, IET does not require us to evaluate the matrix elements in Eq.~(\ref{eq:3}), or to even know the true wave functions~\cite{pra,ann}. From $\xi^{\rm physisorped}_{\rm O_{WAT}-H_{I}} < \xi^{\rm isolated}_{\rm O_{WAT}-H_{I}}$, Eqs.~(\ref{eq:1}) and~(\ref{eq:3}), we can arrive at the conclusion we stated earlier$-$ water molecule physisorption (as depicted in Fig.~\ref{fig:2}B) causes a stronger electric dipole moment between O$_{\rm WAT}$ and H$_{\rm I}$. 

Apart from that, smaller $\xi^{\rm O_{WAT}}_{\rm H_{I}}$ also means a weaker H$_{\rm I}^{\delta +}$$-$O$_{\rm WAT}^{\rm \delta -}$ covalent bond or smaller $\nu_{\rm OH}$. The covalent bonds considered here are of the same kind, $\rm O_{WAT}-H_{I}$ and $\rm O_{WAT}-H_{\rm II}$, hence we need not be concerned with the inverse proportionality between the polarizability and bonding energy discussed in Ref.~\cite{cphc}, in which the inverse proportionality was calculated and justified between two different gas molecules (CO and O$_2$). Interestingly, the measured $\nu_{\rm OH}$ value for a single water molecule adsorbed on MgO monolayer is 448 meV at 4.2 K~\cite{shin}, which is indeed within the expected range ($<$ 454 meV). One should also be aware here that the O$-$H covalent bond in water molecules discussed thus far is still very much stronger than the hydrogen-type bonds, whether the molecule is adsorbed on MgO surface or exists in liquid water or ice.   

We have now established the reasons why and how one of the covalent bond in a single water molecule (H$_{\rm I}^{\delta +}$$-$O$_{\rm WAT}^{\rm \delta -}$: see Fig.~\ref{fig:2}B) can be weakened with physisorption (similar to hydrogen-bond effect). The consequence of this scenario is that one will obtain a stronger H$_{\rm II}^{\delta +}$$-$O$_{\rm WAT}^{\rm \delta -}$ covalent bond as a result of the weaker $\textbf{p}$ between H$_{\rm II}^{\delta +}$ and O$_{\rm WAT}^{\rm \delta -}$ because $^{\rm H_{I}}\delta +$ $<$ $^{\rm H_{II}}\delta +$ and therefore, $\xi^{\rm H_{I}}_{\rm O_{WAT}} < \xi^{\rm H_{II}}_{\rm O_{WAT}}$. Here, H$_{\rm II}^{\delta +}$ is a dangling (not adsorbed) ion. These inequalities tell us that H$_{\rm I}^{\delta +}$$-$O$_{\rm WAT}^{\rm \delta -}$ is the weaker covalent bond, and is susceptible to break in the presence of interlayer tunneling electrons. This bond becomes weaker because the electron, which is originally from H$_{\rm I}$ that was shared with O$_{\rm WAT}$ to form the covalent bond is no longer effectively shared between H$_{\rm I}$ and O$_{\rm WAT}$ in the presence of the above-stated physisorption (or in the presence of hydrogen bonds). This in turn implies that the electron that was originally from H$_{\rm I}$ is now strongly bound to O$_{\rm WAT}$ because H$_{\rm I}$ now interacts attractively with the polarized electrons from O$_{\rm SUR}$. In fact, this conclusion is also in agreement with the experimentally indirectly observed dissociated chemical species (OH$^{-}$) that is found to be adsorbed on a different site (bridge site) after the water molecule dissociation~\cite{shin}. In summary, we have given a rigorous and self-consistent explanation on the water molecule dissociation reported in Ref.~\cite{shin} using IET and its polarizability functional. Along the way, we have also identified which O$-$H covalent bond is weaker or most likely to be breakable, again in agreement with the experiment~\cite{shin}. 

Now we are ready to prove the correctness of the Hermansson blue-shifting hydrogen-bond theory~\cite{her1,her2,her3} in a straightforward manner using Eqs.~(\ref{eq:1}),~(\ref{eq:3}) and our hydrogen-bond-like physisorption scenario between O$_{\rm SUR}$ and H$_{\rm I}$. For example, at intermolecular equilibrium distance, the electric-field interactions and exchange overlap were found to be responsible for the negative value for $\rm d\textbf{p}/d$$r^{\rm H_{I}}_{\rm O_{SUR}}$ that gives rise to the blue-shifting hydrogen bond~\cite{her3}. This effect can be understood within IET by writing

\begin {eqnarray}
\frac{\rm d\textbf{p}}{{\rm d}r^{\rm H_{I}}_{\rm O_{SUR}}} \propto \frac{\rm d\textbf{p}}{\rm d\xi_{\rm O_{SUR}}} = -\lambda\frac{e^2e^{\lambda (E_{\rm F}^0-\xi_{\rm O_{SUR}})}}{M(\omega_{\rm ph}^2 - \omega^2)}\textbf{E}^{\rm H_{I}}_{\rm O_{SUR}}. \label{eq:4}
\end {eqnarray}  

Here, $r^{\rm H_{I}}_{\rm O_{SUR}}$ is the effective separation (distance) between H$_{\rm I}$ and O$_{\rm SUR}$ ions, $\xi_{\rm O_{SUR}}$ denotes the ionization energy that determines the polarizability of O$_{\rm SUR}$ ion and $\textbf{E}^{\rm H_{I}}_{\rm O_{SUR}}$ is the microscopic electric field between O$_{\rm SUR}$ and H$_{\rm I}$ ions. The effective separation here means the distance from the outer core electron (unpolarized) of O$_{\rm SUR}$ ion to the outer radius of H$_{\rm I}$ ion. This definition avoids the contribution from different ionic sizes. Equation~(\ref{eq:4}) clearly points out the existence of blue-shifting attractive interaction (physisorption) for small $\xi_{\rm O_{SUR}}$. This stronger interaction is similar to the blue-shifting hydrogen-bond effect discussed in Ref.~\cite{her3}. The analysis that follows to explain the physisorption-induced bond (similar to hydrogen bonds) is different such that it requires the energy levels from O$_{\rm SUR}$ do not cross (to form covalent or ionic bond) with the energy levels associated to H$_{\rm I}$. This is why $\xi$ is deliberately written as $\xi_{\rm O_{SUR}}$, and not $\xi^{\rm H_{I}}_{\rm O_{SUR}}$, in which, the latter notation is for the bonds with crossed energy-levels between O$_{\rm SUR}$ and H$_{\rm I}$.  

The validity of the above proportionality (Eq.~(\ref{eq:4})) can be verified by noting that for this particular system, $r^{\rm H_{I}}_{\rm O_{SUR}}$ $\propto$ $\xi_{\rm O_{SUR}}$ because smaller $\xi_{\rm O_{SUR}}$ implies stronger Coulomb attraction between (the easily polarizable) O$^{\delta -}_{\rm SUR}$ and H$^{\delta +}_{\rm I}$ ions, and therefore smaller $r^{\rm H_{I}}_{\rm O_{SUR}}$. This means that if H$^{\delta +}_{\rm I}$ ion is positioned far from O$^{\delta -}_{\rm SUR}$ (large $r^{\rm H_{I}}_{\rm O_{SUR}}$), then H$^{\delta +}_{\rm I}$ will induce a small electronic polarization on O$^{\delta -}_{\rm SUR}$, which seems to imply that $\xi_{\rm O_{SUR}}$ is large. We stress here that this implication does not provide a valid cause-and-effect relationship at all, because obviously large $r^{\rm H_{I}}_{\rm O_{SUR}}$ does not cause $\xi_{\rm O_{SUR}}$ to be large (to give small polarization) in any way. However, the opposite is true, meaning, large $\xi_{\rm O_{SUR}}$ causes the effective separation to be large, which explains why $\xi$ can be used to determine the crystal structure of a particular system for a given condition. Recall here that we have applied this proportionality ($r \propto \xi$) earlier such that $r^{\rm MgO}_{\rm Ag}$ $\propto$ $\xi^{\rm MgO}_{\rm Ag}$ in which, large MgO surface polarization (due to small $\xi^{\rm MgO}_{\rm Ag}$) is expected if $r^{\rm MgO}_{\rm Ag}$ is small. Here, small $\xi^{\rm MgO}_{\rm Ag}$ is due to the free-electrons contributed by the metallic Ag substrate.

Next, we need to exclude an alternative water molecule configuration (Fig.~\ref{fig:2}A) proposed earlier based on the experimental results obtained by Shin et al.~\cite{shin}. In this configuration, the negatively charged O$_{\rm WAT}^{\rm \delta -}$ ion at the bridge site, remains there after the dissociation, which does not require any re-adsorption of the dissociated product (OH$^{-}$). Whereas, the most stable configuration depicted in Fig.~\ref{fig:2}B requires re-adsorption of OH$^{-}$ on the bridge site after dissociation. The configuration in Fig.~\ref{fig:2}A requires the physisorped O$^{\delta -}_{\rm WAT}$ ion onto the bridge site to satisfy the inequality $V_{\rm Coulomb}^{\rm attract} > V_{\rm Coulomb}^{\rm repel}$, hence there is an effective attraction between O$^{\delta -}_{\rm WAT}$ ion and its nearest neighbors (two Mg$^{2+}$ and two O$^{\delta -}_{\rm SUR}$ ions). Such an attractive interaction is energetically favorable (in nonequilibrium conditions: fluctuating energy at nanoscale) because it is similar to the experimentally observed OH$^-$ ion adsorbed on the bridge site after dissociation. The oxygen ion from a single water molecule physisorped on the bridge site will lead to the inequality $^{\rm O_{WAT}}_{\rm physisorped}\delta -$ $<$ $^{\rm O_{WAT}}_{\rm isolated}\delta -$ due to the existence of an effective $V_{\rm Coulomb}^{\rm attract}$ between O$^{\delta -}_{\rm WAT}$ and Mg$^{2+}$. 

Contrary to the discussion given earlier based on Fig.~\ref{fig:2}B, the configuration depicted in Fig.~\ref{fig:2}A gives rise to a stronger $e$-$e$ interaction within the O$^{\delta -}_{\rm WAT}$ ion if $V_{\rm Coulomb}^{\rm attract}$ between O$^{\delta -}_{\rm WAT}$ and Mg$^{2+}$ ions gets stronger. For example, the $e$-$e$ interaction stated here is between the two electrons from H$_{\rm I}$ and H$_{\rm II}$ ions. As a consequence, the stronger $V_{\rm Coulomb}^{\rm attract}$ reduces the screening strength between these two electrons (now strongly interacting) within the physisorped O$^{\delta -}_{\rm WAT}$ ion that allows us to write $V_{\rm Coulomb}^{\rm attract}$ $\propto V_{\rm sc}^{\rm O_{WAT}}$. As such, from Eqs.~(\ref{eq:1}) and~(\ref{eq:2}), we can readily conclude that one needs a higher energy to excite an electron from the physisorped O$^{\delta -}_{\rm WAT}$ ion (due to $\xi^{\rm isolated}_{\rm O_{WAT}} < \xi^{\rm physisorped}_{\rm O_{WAT}}$), which implies stronger covalent bonds between O$^{\delta -}_{\rm WAT}$ and its two dangling hydrogen ions. This means that the asymmetric vibrational frequency, $\nu_{\rm OH}$ is expected to be larger than that of an isolated water molecule, or $\nu_{\rm OH} >$ 454 meV (water vapour), which is invalid based on the experimental result (448 meV) reported in Ref.~\cite{shin}.             

\begin{figure}
\begin{center}
\scalebox{0.6}{\includegraphics{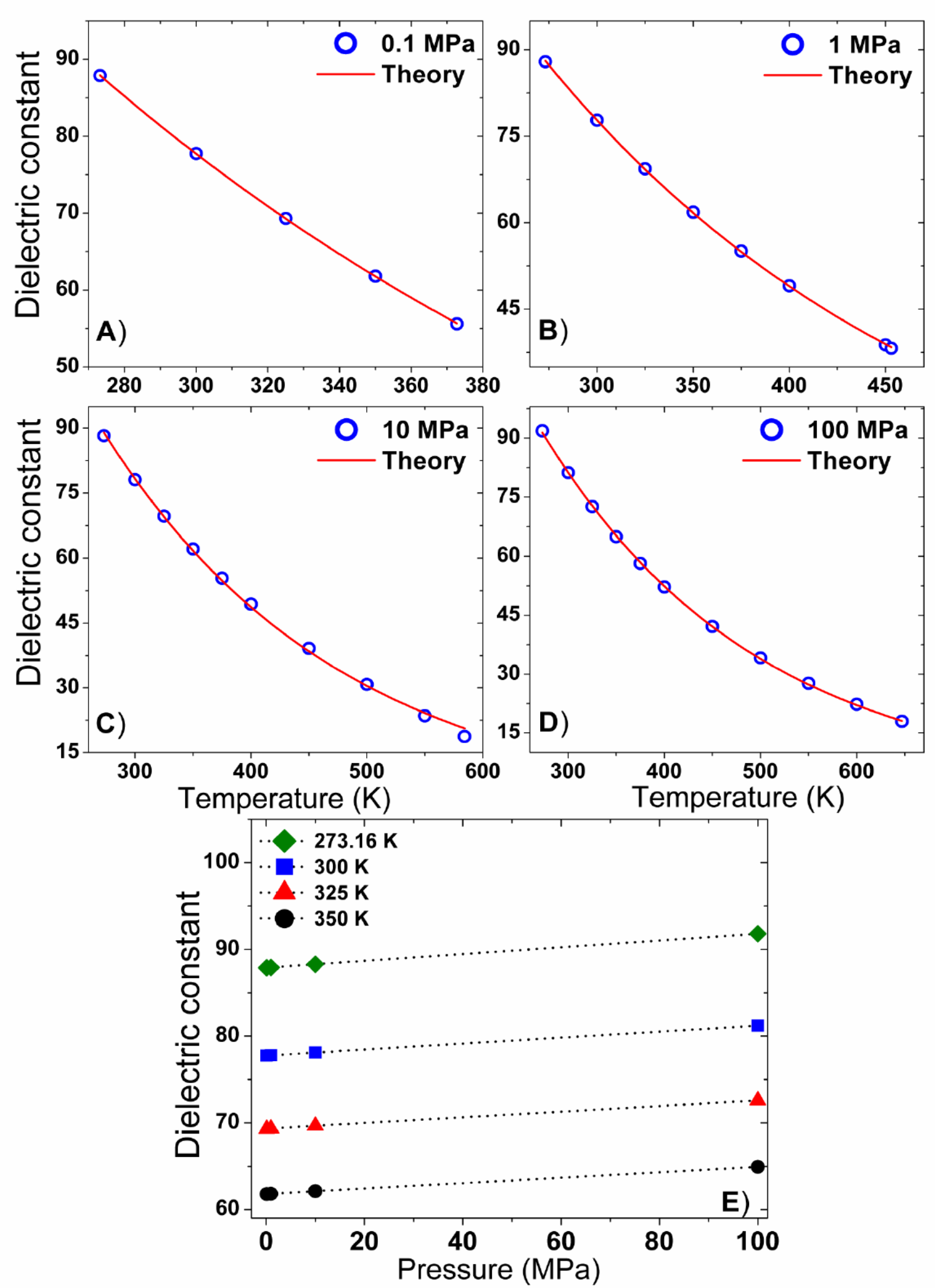}}
\caption{Temperature-dependent static dielectric constant data of water for different pressures in MPa (A) 0.1, (B) 1, (C) 10, and (D) 100. The solid lines in (A)$-$(D) follow Eq.~(\ref{eq:6}) with $303.2 < \gamma < 337.1$, $10/2254 < \eta < 10/2042$, and $O$[$T,P,\rho$] = 0. (E) Shows small linear-increment to the static dielectric constant of water when the pressure gets larger from 0.1 to 100 MPa, for a given temperature. These linear dependents are indicated with dotted lines. All the experimental data points were obtained from Ref.~\cite{lide}.}
\label{fig:3}
\end{center}
\end{figure}

There is however one last obstacle that needs to be overcome in order to complete the analysis on water-molecule splitting mechanism unequivocally, and with high-level self-consistency$-$ the $T$-dependent static dielectric constant ($\epsilon$) of water must obey the phenomenological theory of dielectricity derived in Ref.~\cite{pla2}. However, the renormalized~\cite{pla2} 

\begin {eqnarray}
\epsilon = 1 + \frac{K_s^2}{|\textbf{k}|^2}\exp\big[{\lambda(E_{\rm F}^0 - \xi)}\big] \label{eq:5}
\end {eqnarray}  

is $T$-independent, where $|\textbf{k}| = k$, $k$ denotes the wave number, $K_s = 3n_0/2\epsilon_0 E_{\rm F}^0e^2$ and $n_0$ is the free-electron carrier density at $T = 0$. Hence, one needs to transform Eq.~(\ref{eq:5}) to obtain a new $T$-dependent constant,

\begin {eqnarray}
\epsilon = 1 + \gamma\exp\big[{-\lambda \eta' k_BT}\big] + O[T,P,\rho]. \label{eq:6}
\end {eqnarray}  

Here, $\gamma$ and $\eta'$ are constants of proportionality, while $O$[$T,P,\rho$] is introduced by-hand to represent other (including higher order) terms that captures the power-law behaviors due to $T$, pressure ($P$) and density-of-water ($\rho$) effects. Traditionally, $\epsilon$ is written as a function that obeys only the power-law with up to ten numerical (both positive and negative) constants~\cite{ue}. This explains why the term $O$[$T,P,\rho$] exists in Eq.~(\ref{eq:6}), and its experimental justifications were also given by Uematsu and Franck~\cite{ue}. 

The above transformation reads $(K_s^2/k^2)\exp{[\lambda(E_{\rm F}^0 - \xi)]} \longrightarrow \gamma\exp{[-\lambda\eta' k_BT]}$, where the physically relevant transformation is $\xi \longrightarrow \eta T$, which can be understood within the water molecule dissociation mechanism explained earlier. This means that we need to establish $\xi \propto T$, which is not a cause-and-effect relationship but can be exploited to carry out the above transformation. The logic behind this proportionality is due to (i) $1/\xi_{\rm H^{+}} \propto (^{\rm O_{WAT}}\delta -, ^{\rm H_{I}}\delta +)$ and (ii) $1/T \propto (^{\rm O_{WAT}}\delta -, ^{\rm H_{I}}\delta +)$, and therefore, $1/\xi_{\rm H^{+}} \propto 1/T$. Most importantly, Eq.~(\ref{eq:6}) also defines the existence of a first-order transition temperature ($T_{\rm FO}$) such that $\gamma\exp[{-\lambda \eta' k_BT}]$ dominates for $T < T_{\rm FO}$ and $O[T,P,\rho] \rightarrow 0$. On the other hand, $O[T,P,\rho]$ becomes the dominant term for $T > T_{\rm FO}$ and $\gamma\exp[{-\lambda \eta' k_BT}] \rightarrow 0$. 

Interestingly, the existence of $T_{\rm FO}$ has been established experimentally and via simulations in Refs.~\cite{kum,xu,liu,kum2}, for instance, the transition from Arrhenius (exponential-type) behavior at low temperatures (below the Widom temperature, $T_{\rm W}$) to the power-law behavior for $T > T_{\rm W}$. However, $T_{\rm FO} \neq T_{\rm W}$ because $T_{\rm FO}$ here means the electronic-phase transition temperature where $T > T_{\rm FO}$ is the regime with no electronic interaction (that can cause changes to the molecular energy levels) between water molecules. Figure~\ref{fig:3} shows the theoretical fits of $\epsilon$ based on Eq.~(\ref{eq:6}) and the range of its fitting parameters, indicating the dominant roles played by the polarization induced $e$-$e$ interaction and $\textbf{p}$ below $T_{\rm FO}$, in agreement with the water-molecule splitting mechanism presented earlier. The correctness of the exponential term is also logically justified because $T$ influences the electronic excitation probability, as it should be. On the contrary, $P$ does not influence the above probability directly nor significantly and therefore, one obtains an almost perfect linear fit between $\epsilon$ and $P$ for a given $T$ (Fig.~\ref{fig:3}E). 

In conclusion, hydrogen produced from the photocatalytic splitting of water may or may not completely replace the polluting fossil and/or the radioactive nuclear fuels because such a replacement depends on the toxicity and recoverability of the catalysts used. But any form of alternative clean energy, including hydrogen from water is worth exploring to diversify the energy supply to some extent. In this regard, we have clearly explained that the root cause responsible to the water molecule splitting is because one of its O$-$H bonds become weaker when exposed to external disturbances. This weaker bond has its hydrogen ion adsorbed on the oxide surface. Such an adsorption exists due to the polarization-induced attractive interaction between a positively charged hydrogen ion and the negatively charged surface oxygen ion. This Coulomb interaction results in the modification of the electron-electron interaction and subsequently to a stronger electric dipole moment, and finally to a weaker O$-$H bond (the hydrogen ion here is adsorbed onto the surface oxygen ion). The other O$-$H bond dangles upward (this hydrogen ion remains unadsorbed). Therefore, this weaker bond, which is susceptible, gives in to the interlayer tunneling electrons and leaves behind the dissociated product, OH$^-$ ion re-adsorbed on the bridge site (surrounded by two magnesium and two oxygen ions). 

On the way to make the above mechanism explicit and unequivocal, we have also excluded one other water molecule configuration because it may be energetically favorable in nonequilibrium conditions for two reasons$-$ (a) this new physisorption does not require re-adsorption of OH$^-$ ion on the bridge site after dissociation and (b) OH$^-$ ion is found to be stable on the bridge site. However, this configuration is found to be invalid due to its larger O$-$H asymmetric vibrational frequency, which is predicted to be larger than that of an isolated water molecule. Apart from that, the water splitting mechanism proven here is also found to be consistent with other established theories, simulations and measurements. In particular, the Hermansson blue-shifting hydrogen-bond theory, the phenomenological theory of dielectricity, and the temperature-dependent static dielectric constant measurements of water for different pressures and below the first-order electronic-phase transition temperature. As a matter of fact, this mechanism will be further exploited to understand the reasons why and how water molecules may activate the large biomolecules, including DNA, and how they bring biological systems to ``life''. 

\section*{Acknowledgments}

This work was supported by Arokia Das Anthony and Cecily Arokiam, as well as the Ad-Futura visiting researcher fellowship (Slovenia) between Feb/2010 and Feb/2011. A.D.A. reserves special thanks to Naresh Kumar Mani for explaining the importance of polar molecules and water in DNA folding mechanism. Z.K. thanks the National Foundation for Science and the Higher Education and Technological Development of the Republic of Croatia. K.E. and M.M. thank Matev$\check{\rm z}$ Pi$\check{\rm c}$man and Rok Zaplotnik for their contributions to develop several animations and figures, respectively, and the Slovenian Ministry of Higher Education, Science and Technology for the scholarships. U.S.S. is grateful to G. Subramani and S. Mekala for the financial support. We are grateful to the anonymous referees for cross-checking all the technical details, and for their comments and suggestions given in Refs.~\cite{referee1,referee2}.


\begin{thebibliography}{plain}
\bibitem{referee1} One can experimentally cross-check the above theoretical claim via the XPS measurements, in which, the binding energy of an oxygen ion adsorbed on a single layer MgO/Ag system can be compared to that of an oxygen ion adsorbed on a multi-layered MgO/Ag system.
\bibitem{referee2} It is also possible for one to cross-check this weakened covalent bond via the appropriate $^2$H solid state NMR experiments. In particular, (i) for a single and free heavy water ($^2$H$_2$O) molecule and (ii) for a $^2$H$_2$O molecule adsorbed on the MgO/Ag system.
\end{thebibliography}
\end{document}